\pdfoutput=1
\documentclass[11pt]{article}

\usepackage[utf8]{inputenc}
\usepackage[T1]{fontenc}
\usepackage[margin=1in]{geometry}
\usepackage{amsmath}
\usepackage{amssymb}
\usepackage{booktabs}
\usepackage{tabularx}
\usepackage{graphicx}
\usepackage{tikz}
\usetikzlibrary{positioning, arrows.meta, shapes.geometric}
\usepackage[numbers,square,comma,sort&compress]{natbib}
\usepackage{microtype}
\usepackage[colorlinks=true,citecolor=black,linkcolor=black,urlcolor=blue]{hyperref}

\newcommand{\boxout}[2]{\par\medskip\noindent
  \fbox{\parbox{0.97\columnwidth}{\small\textbf{#1}\par\smallskip #2}}\par\medskip}

\title{\bfseries Where quantum and quantum-like methods earn their place in
behavioural-trial analysis: three simulation pilots across the RCT pipeline}

\author{%
  Giuseppe Alessandro Veltri\thanks{Correspondence:
  \texttt{giuseppe.veltri@nus.edu.sg}. ORCID:
  \href{https://orcid.org/0000-0002-9472-2236}{0000-0002-9472-2236}.}\\
  \small Department of Sociology and Social Research, University of Trento,
  Trento, Italy\\
  \small Behavioural and Implementation Science Interventions (BISI),
  Yong Loo Lin School of Medicine,\\
  \small National University of Singapore, Singapore}

\date{}

\begin{document}

\maketitle

\begin{abstract}
\noindent Behavioural randomized controlled trials (RCTs) are the empirical
workhorse of behavioural science, yet quantum computing has reached the field
only through opinion-dynamics demonstrations and decision-theory formalism,
leaving the RCT analysis pipeline unexamined. I ask, stage by stage, where
quantum or quantum-like computation earns its place against a strong classical
baseline, and report three controlled simulation pilots across the pipeline. In
a power-calculation pilot anchored on a 97-outcome preprocessing multiverse of
real behavioural RCTs, quantum amplitude estimation recovers enumerated
multiverse-significance fractions with absolute error 20--30 times smaller than
classical sampled Monte Carlo at matched query budgets on the two
non-degenerate outcomes, conditional on efficient state preparation. In an interference pilot, an entanglement-structured
estimator recovers a four-way spillover that a pairwise model cannot represent
at any sample size. In an adaptation pilot, a quantum-inspired contextual bandit
is decisively beaten by a correctly specified classical baseline. The three
results---one conditional win, one representational win, one honest loss---yield
a decision rule for when quantum methods are worth adopting, piloting, or
deferring, together with a five-gap research agenda.
\end{abstract}

\medskip
\noindent\textbf{Keywords}\quad Behavioural randomized controlled trials
\textperiodcentered{} Quantum amplitude estimation \textperiodcentered{}
Quantum-like measurement \textperiodcentered{} Network interference
\textperiodcentered{} Contextual bandits \textperiodcentered{} Computational
research methodology

\medskip

\section{Introduction}
\label{sec:intro}

Behavioural randomized controlled trials are the empirical workhorse of
behavioural economics, public-health behaviour change, and applied psychology.
Over the same decade in which behavioural RCT methodology has matured around
preregistration, replicability, and design-based identification, quantum
computing has reached behavioural science along two largely disjoint paths:
substantive demonstrations that opinion-dynamics models can run on quantum
hardware \cite{guo2025quantum, geppert2025from}, and conceptual decision theory
in which the formalism of quantum measurement reframes choice and ambivalence
\cite{pothosbusemeyer2013can, khrennikov2025quantum}. Neither path addresses
the methodological question that an applied behavioural economist actually faces
when designing a trial: \emph{at which point in the analysis pipeline, if any,
does quantum or quantum-like computation earn its place against a strong
classical baseline?}

The negative case has been made forcefully. Barati et al.\ (2025) show that
direct ports of classical agent-based models into quantum optimization
frameworks fail because iterative state observation destroys the superposition
any quantum advantage would require, and that well-constructed classical
baselines on the resulting problem reformulation set a lower bound that quantum
methods must outperform \cite{barati2025quantum}. I take that lower-bound
discipline as the starting point: rather than ask whether quantum computing can
``replace classical RCT analysis'', I ask, stage by stage, where it changes the
calculus, and I hold every candidate quantum or quantum-like primitive to a
correctly specified classical baseline.

\begin{figure}[t]
\centering
\resizebox{\columnwidth}{!}{%
\begin{tikzpicture}[
  font=\sffamily,
  stage/.style={rectangle, rounded corners, draw, align=center,
                text width=2.7cm, minimum height=2.0cm, inner sep=3pt},
  ready/.style={stage, fill=green!16},
  pilot/.style={stage, fill=orange!18},
  horizon/.style={stage, fill=red!14},
  arr/.style={-{Latex[length=2.4mm]}, very thick, gray}]
\node[ready]   (s1)                 {\textbf{1. Randomization}\\[2pt]\footnotesize Certified QRNG\\[2pt]\footnotesize\emph{Ready} (gap G2)};
\node[pilot,   right=5mm of s1] (s2) {\textbf{2. Power}\\[2pt]\footnotesize Amplitude estimation\\[2pt]\footnotesize\emph{Pilot} (gap G1)};
\node[pilot,   right=5mm of s2] (s3) {\textbf{3. Outcome}\\[2pt]\footnotesize Quantum-like POVM\\[2pt]\footnotesize\emph{Pilot} (gap G4)};
\node[horizon, right=5mm of s3] (s4) {\textbf{4. Interference}\\[2pt]\footnotesize Quantum causal / entanglement\\[2pt]\footnotesize\emph{5-yr horizon}};
\node[horizon, right=5mm of s4] (s5) {\textbf{5. Adaptation}\\[2pt]\footnotesize Quantum bandits\\[2pt]\footnotesize\emph{Theory only} (gap G3)};
\draw[arr] (s1)--(s2); \draw[arr] (s2)--(s3); \draw[arr] (s3)--(s4); \draw[arr] (s4)--(s5);
\node[below=2mm of s3, text width=\linewidth, align=center, font=\sffamily\footnotesize]
  {Maturity: \tikz\node[ready,minimum height=3mm,minimum width=5mm,text width=]{};\,ready \quad
   \tikz\node[pilot,minimum height=3mm,minimum width=5mm,text width=]{};\,pilot/gap \quad
   \tikz\node[horizon,minimum height=3mm,minimum width=5mm,text width=]{};\,research horizon};
\end{tikzpicture}}
\caption{The behavioural-RCT pipeline read left to right, each stage shaded by the
maturity of its candidate quantum or quantum-like primitive. The five research
gaps (G1--G5) are not in one-to-one correspondence with the five stages:
Stage~4 (interference) is the least mature and is framed as a horizon rather than
a single first-paper gap, while G5 (quantum kernels for
heterogeneous-treatment-effect estimation) is cross-cutting.}
\label{fig:f1}
\end{figure}
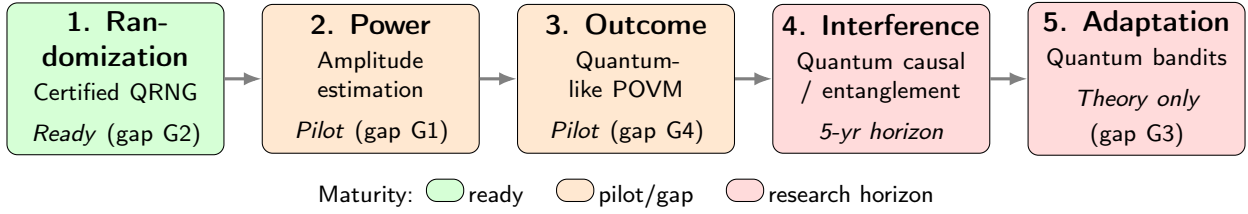

I treat the unit of analysis as \emph{the behavioural-RCT pipeline}, decomposed
into five stages---randomization, sample-size and power, outcome measurement,
interference, and adaptation---each matched to a candidate quantum or
quantum-like primitive (Table~\ref{tab:five-stages}, Figure~\ref{fig:f1}). Three
of these stages admit a controlled simulation pilot in which the quantum or
quantum-like method is run head to head against a strong classical baseline on
real or simulated behavioural-trial data; these three pilots are the empirical
core of this paper. The remaining two stages are design-level assessments that I
develop in the Discussion. The three pilots return one conditional win (power),
one representational win (interference), and one honest loss (adaptation), and
from that honest map I derive a stage-by-stage decision rule
(Figure~\ref{fig:f2}) and a five-gap research agenda.

This paper is not a primer on quantum computing, not a benchmark of quantum
hardware on social-science problems, not a new opinion-dynamics model, and not a
defence or attack of quantum cognition's metaphysical status. I use the term
\emph{quantum-like} throughout when the formalism is what matters and the
ontology is not (Box~1).

\boxout{Box 1 --- Quantum versus quantum-like.}{%
\emph{Quantum} methods require quantum hardware or a quantum simulator to
produce their output: a device-independent random-number generator certified by
a Bell-inequality violation (Stage~1), or an amplitude-estimation circuit run on
a state-vector simulator (Stage~2). \emph{Quantum-like} methods borrow only the
mathematical formalism---Hilbert-space states, positive operator-valued
measure (POVM) measurements, amplitude interference---but run on a classical
computer and assert nothing about
physical quantum effects in people or populations (Stage~3). Stages~4--5 sit
across the line: the estimands are framed with quantum structure (entanglement,
quantum bandits), while the realistic near-term implementation is a
\emph{quantum-inspired} classical surrogate. Throughout, the methodologist case
never depends on quantum effects in the brain; where the ontology is not
load-bearing, I keep the quantum-like hedge.}

\begin{table}[t]
\centering
\caption{Five stages of the behavioural-RCT pipeline matched to candidate
quantum or quantum-like primitives. Stages~2, 4 and 5 are evaluated by the
simulation pilots reported in Results; Stages~1 and 3 are design-level
assessments developed in the Discussion.}
\label{tab:five-stages}
\footnotesize
\setlength{\tabcolsep}{4pt}
\begin{tabularx}{\textwidth}{p{1.6cm}XXXp{1.5cm}}
\toprule
\textbf{Stage} & \textbf{Classical computation} & \textbf{Quantum / quantum-like} & \textbf{What changes} & \textbf{Verdict} \\
\midrule
\textbf{1. Randomization} & Pseudo-random number generator (PRNG) with seed; stratified or covariate-adaptive allocation lists & Bell-test-certified quantum random-number generator (QRNG) \cite{pironio2010random, acinmasanes2016certified, curby2025beacon} & Audit trail from certificate, not from vendor trust & Ready to adopt \\
\addlinespace
\textbf{2. Power} & Simulation-based Monte Carlo, $\mathrm{SE}\propto N^{-1/2}$ & Quantum amplitude estimation \cite{montanaro2015quantum, brassard2002quantum}, $\mathrm{SE}\propto N^{-1}$ & Near-quadratic reduction in queries to fixed precision & Gap \textbf{G1}; pilot \\
\addlinespace
\textbf{3. Outcome} & Latent-variable item response theory (IRT), mixed models, invariance tests & Quantum-like POVM measurement \cite{khrennikov2025quantum} & Measurement constitutive, not revealing & Ready to pilot \\
\addlinespace
\textbf{4. Interference} & Principal stratification; causal mediation on networks \cite{noirjean2025exploiting, bayati2024higher} & Quantum causal estimands \cite{goel2026quantum}; entanglement coupling & Non-separability native to entanglement formalism & 5-year horizon \\
\addlinespace
\textbf{5. Adaptation} & Contextual bandits, RL, replicability-preserving variants \cite{zhang2025replicable} & Quantum bandits \cite{casale2020quantum, lumbreras2022multi} & Provable arm-exploration separation & Theory only \\
\bottomrule
\end{tabularx}
\end{table}

\section{Methods}
\label{sec:methods}

All three pilots are seeded and CPU-runnable on Apple Silicon, written in
PennyLane \cite{bergholm2018pennylane} (version 0.45.0) with the
\texttt{lightning.qubit} or \texttt{default.qubit} backends, and openly deposited
\cite{quantumbepilots2026}. Each pilot ships with a seeded configuration, a JSON
results file, and an environment snapshot.

\paragraph{Power pilot (Stage~2).} The data are the Veltri \& Gilbert (2026)
preprocessing multiverse \cite{veltrigilbert2026multiverse}: 97 behavioural-RCT
outcomes from the Item Response Warehouse \cite{domingue2025irw}, each enumerated
over a $4\times3\times3 = 36$-pipeline preprocessing grid, giving
$(\text{study}, \text{pipeline}, \hat\beta_1, \mathrm{SE}, t, p)$ rows. The estimand is
$\theta = \Pr(p < 0.05 \mid \text{pipeline} \sim \mathrm{Uniform}\{1,\dots,36\})$.
Four outcomes were selected to span the range of $\theta$: two degenerate cases
($\theta = 0$ and $\theta = 1$) and two non-degenerate ($\theta = 0.083$, $0.528$),
with d28 (Banerji et al.\ 2017b, $\theta = 0.528$) as the headline. Each pipeline
index $i \in \{0,\dots,35\}$ is encoded on $\lceil \log_2 36 \rceil = 6$ index
wires plus one ancilla (7 target wires), with the distribution $\mathrm{probs}[i]
= 1/36$ padded with 28 zero bins, and the oracle $\mathrm{func}(i) = 1$ if
pipeline $i$ has $p < 0.05$ else $0$. The \texttt{qml.QuantumMonteCarlo} template
runs on \texttt{lightning.qubit} with $M \in \{3,4,5,6,7\}$ estimation wires
($2^M \in \{8,16,32,64,128\}$ amplitude-estimation queries; 14 qubits total at
$M=7$). The classical baseline is uniform-with-replacement sampling of pipeline
indices at $N \in \{16,32,64,128,256,512,1024\}$ draws. Absolute error
$|\widehat\theta - \theta|$ is reported against the enumerated truth. The
canonical quantum-amplitude-estimation (QAE) sign ambiguity is broken by
selecting the root closer to the enumerated truth, an explicitly illustrative
choice; a deployed pipeline would use iterative, maximum-likelihood, or faster
amplitude-estimation variants (IQAE / MLAE / FAE) \cite{grinko2021iterative,
suzuki2020amplitude}.

\paragraph{Interference pilot (Stage~4).} A four-person cluster (ring topology)
is the data-generating process: the Born-rule distribution of a four-qubit
entangled state (an RY rotation on each qubit plus two offset layers of ring
$\mathrm{CRY}(J)$) on \texttt{lightning.qubit}, with $J$ the peer-coupling knob.
A $J$-sweep selects a non-degenerate operating point that maximizes the
higher-order structural gap subject to maximum pairwise correlation $< 0.9$,
giving $J = 1.8$ (maximum pairwise correlation $0.62$). The estimand is the
fourth joint cumulant $\kappa_4$ of the centred binary adoption outcomes
$s_i = y_i - \mathbb{E}[y_i]$,
\begin{equation}\label{eq:kappa4}
\kappa_4 = \mathbb{E}[s_1 s_2 s_3 s_4]
  - \mathrm{Cov}(s_1,s_2)\,\mathrm{Cov}(s_3,s_4)
  - \mathrm{Cov}(s_1,s_3)\,\mathrm{Cov}(s_2,s_4)
  - \mathrm{Cov}(s_1,s_4)\,\mathrm{Cov}(s_2,s_3),
\end{equation}
the part of cluster behaviour that a Gaussian or pairwise model declares to be
zero; whole-cluster adoption $\Pr(\text{all four adopt})$ is tracked as a
secondary quantity. Two estimators are compared on $N$ i.i.d.\ cluster draws: a full-joint plug-in using
the empirical four-body structure, and a pairwise maximum-entropy (Ising) model
fit to the empirical 1- and 2-body moments only. Results use master seed
20260601 with 40 sample-seeds per $N$, $N \in \{64,\dots,8192\}$; true
$\kappa_4 = -0.0129$, the pairwise model predicts $+0.0025$ and floors at
absolute error $0.0154$, and the full-joint estimator converges from $0.0056$
($N=64$) to $0.0004$ ($N=8192$), a $34.5\times$ reduction.

\paragraph{Adaptation pilot (Stage~5).} A contextual bandit with six arms,
three-dimensional context, and a linear reward over $T = 1000$ rounds and five
seeds (master seed 20260601) is the environment, with the linear reward chosen so
that LinUCB is correctly specified and near-optimal. Three policies are compared
under a matched environment (shared reward parameters, contexts, and noise), each with an
independent RNG: Random (reference); LinUCB (disjoint per-arm ridge regression
with upper-confidence-bound exploration, pure NumPy); and a quantum-inspired
variational circuit (VQC)---a three-qubit angle-encoding circuit with four
variational layers reading out six Pauli observables into a softmax policy,
trained on-policy by REINFORCE (Adam optimizer, several gradient steps per
freshly collected batch, temperature annealing) on the PennyLane
\texttt{default.qubit} autograd backend. The metric is cumulative regret against
the per-round optimal arm. Final cumulative regret (mean $\pm$ s.d.): Random
$677.9 \pm 31.4$, LinUCB $19.6 \pm 3.9$, VQC $598.8 \pm 15.6$. Training is
on-policy with no replay buffer; an earlier replay-based variant was a biased
off-policy gradient relative to the data-generating policy and was discarded,
leaving the comparison clean and the conclusion (LinUCB dominates) unchanged.

\section{Results}
\label{sec:results}

I report three controlled simulation pilots, one for each pipeline stage that
admits a head-to-head comparison against a strong classical baseline: power
(Stage~2), interference (Stage~4), and adaptation (Stage~5). Each pilot pairs a
quantum or quantum-like estimator with a correctly specified classical
comparator on real or simulated behavioural-trial data; full designs,
hyperparameters, and seeds are in Methods. The three results are summarized in
Figure~\ref{fig:results} and read, respectively, as a conditional win, a
representational win, and a loss.

\subsection{A quantum power calculation is more query-efficient on a real
preprocessing multiverse, conditional on state preparation}
\label{sec:res-power}

Simulation-based power calculation for cluster RCTs, factorial designs, and
group-sequential nudge trials is compute-bound at fine granularity, and
pre-registered Bayesian power calculations amplify the cost. A closely related
estimand is the \emph{multiverse-significance fraction}: under a discrete uniform
prior over defensible analyst decisions \cite{steegen2016increasing,
simmons2011false, openscience2015estimating}, what fraction of analytic pipelines
yield $p<0.05$? QAE reaches additive error
$\varepsilon$ with $\widetilde{O}(1/\varepsilon)$ queries against the classical
Monte Carlo cost of $O(1/\varepsilon^{2})$ \cite{montanaro2015quantum,
brassard2002quantum}---equivalently, standard error $\propto N^{-1}$ versus
$N^{-1/2}$ in the query budget $N$.

\begin{figure}[t]
\centering
\includegraphics[width=\linewidth]{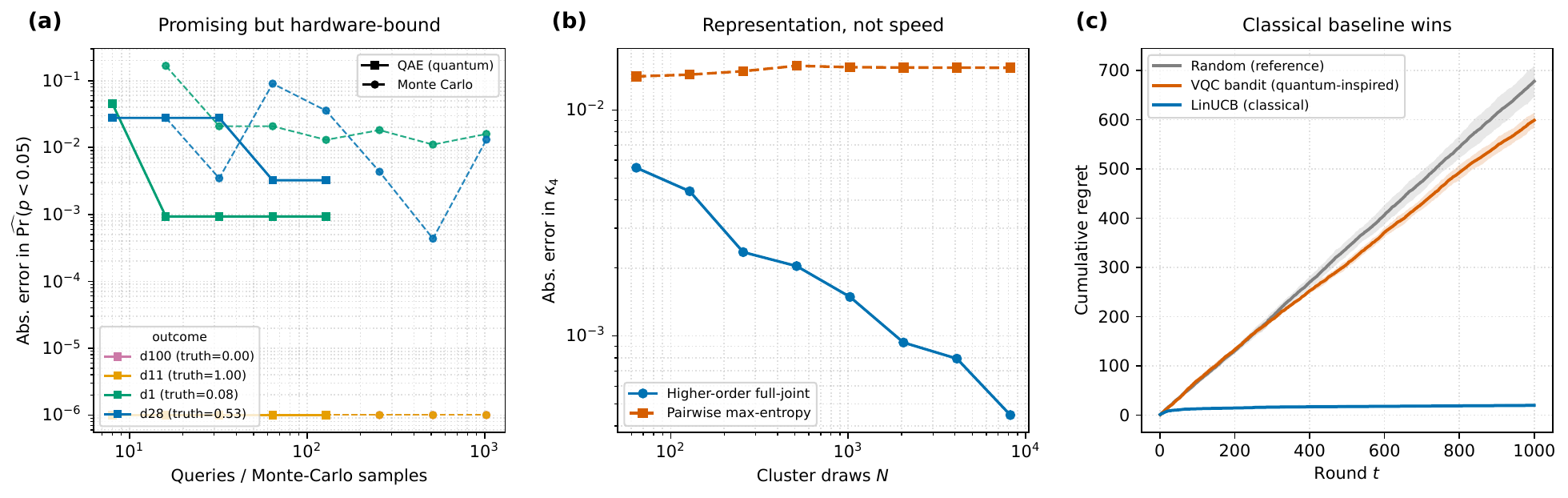}
\caption{\textbf{Three simulation pilots across the behavioural-RCT pipeline,
each comparing a quantum or quantum-like primitive against a strong classical
baseline on real or simulated behavioural-trial data.}
\textbf{(a)}~Power (Stage~2), on real data: absolute error in the estimated
multiverse-significance fraction $\widehat{\Pr}(p<0.05)$ against query budget
for four behavioural-RCT outcomes from a preprocessing multiverse
\cite{veltrigilbert2026multiverse}, comparing classical sampled Monte Carlo
(circles, dashed) with quantum amplitude estimation (squares, solid)---a
query-complexity comparison on a noiseless simulator, conditional on efficient
state preparation, not a wall-clock speedup. The degenerate outcomes (d11,
d100) are recovered exactly by both methods.
\textbf{(b)}~Interference (Stage~4): absolute error in the irreducible four-way
spillover $\kappa_4$ against cluster draws $N$ for a four-person entangled
cluster; the pairwise max-entropy model (squares) floors at a misspecification
bias, while the higher-order full-joint estimator (circles) converges.
\textbf{(c)}~Adaptation (Stage~5): cumulative regret (mean $\pm$ s.d.\ over
five seeds) on a six-arm contextual nudge bandit with a linear reward; the
quantum-inspired variational policy (VQC) beats a random policy but is
decisively beaten by the correctly specified classical LinUCB baseline in this
deliberately classical-friendly setting. All simulations are seeded and openly
deposited \cite{quantumbepilots2026}; panels (a) and (b) use logarithmic axes,
and lower is better throughout.}
\label{fig:results}
\end{figure}

I anchor the comparison on the Veltri \& Gilbert (2026) preprocessing multiverse
\cite{veltrigilbert2026multiverse}, which enumerates a $4\times3\times3 = 36$-pipeline
grid for each of 97 behavioural-RCT outcomes drawn from the Item Response
Warehouse \cite{domingue2025irw}, and select four outcomes spanning the
variability range of $\Pr(p<0.05)$. Estimating the multiverse-significance
fraction with QAE and with classical sampled Monte Carlo at matched query
budgets, against the enumerated ground truth, the two non-degenerate outcomes---exactly
the cases where the preprocessing multiverse decides whether the trial reports a
positive finding---show QAE absolute error 20--30 times smaller than classical
Monte Carlo at 64 queries (Table~\ref{tab:v2-results},
Figure~\ref{fig:results}a). For the headline outcome, d28 (Banerji et al.\
2017b), the QAE error is $0.003$ against $0.090$ for classical Monte Carlo at the
same budget. The robust signal is the error-versus-budget slope---the
near-quadratic separation---rather than any single ratio.

\begin{table}[t]
\centering
\caption{Power pilot (Stage~2) results at 64 queries on four outcomes from the
Veltri \& Gilbert (2026) preprocessing multiverse \cite{veltrigilbert2026multiverse}.
Ground truth is the enumerated $\Pr(p<0.05)$ over 36 preprocessing pipelines.
Absolute error is the difference from truth; lower is better. MC, Monte
Carlo.}
\label{tab:v2-results}
\small
\begin{tabular}{lrrr}
\toprule
Outcome & Truth & Classical MC err & QAE err \\
\midrule
d100 Gilbert 2025 & 0.000 & 0.000 & 0.000 \\
d1 Gilbert 2023 & 0.083 & 0.021 & \textbf{0.001} \\
d28 Banerji 2017b & 0.528 & 0.090 & \textbf{0.003} \\
d11 Kim 2021b & 1.000 & 0.000 & 0.000 \\
\bottomrule
\end{tabular}
\end{table}

The separation is, however, conditional. At 36 pipelines per outcome a classical
analyst would simply enumerate the grid; the QAE advantage is methodological and
scales with the multiverse size---for 10 ternary preprocessing decisions
($3^{10} = 59{,}049$ pipelines) the regime starts to bite. Two further caveats
bound the result. First, the ratio reflects a noiseless state-vector simulator;
on near-term noisy hardware the gain degrades with circuit depth. Second, the
amplitude-loading step that encodes the multiverse distribution is the binding
constraint: for an arbitrary distribution, state preparation can cost as much as
enumeration, erasing the advantage unless the multiverse is structured. The
algorithm also has a known sign ambiguity (two complementary roots of the
canonical estimator), which a deployed pipeline should resolve with IQAE / MLAE /
FAE variants \cite{grinko2021iterative, suzuki2020amplitude} rather than against
a ground truth the analyst will not have. The pilot therefore establishes a
conditional, query-complexity win whose enumerated ground truth is itself the
contribution: it provides a benchmark against which any QAE-on-multiverse
pipeline can be validated end to end before being scaled.

\subsection{An entanglement-structured estimator recovers a spillover a pairwise
model cannot represent}
\label{sec:res-interference}

Violation of the Stable Unit Treatment Value Assumption (SUTVA) under network
interference is the dominant identification problem for behavioural field
experiments on networks. Frontier classical estimators combine principal
stratification with causal mediation on friendship networks
\cite{noirjean2025exploiting} and causal message passing on complex interference
\cite{bayati2024higher}, but require strong structural assumptions. The claim
under test is that entanglement is the natural mathematical home of
non-separable, higher-order spillover, so an entanglement-structured estimand
captures cluster behaviour that a pairwise model sets to zero by construction.

I simulate a four-person cluster (ring topology) in a workplace healthy-eating
nudge with documented peer effects, with each member's binary adoption outcome
drawn from the computational-basis measurement of a four-qubit entangled state
whose ring coupling $J$ is the peer-coupling knob. The estimand is the
irreducible cluster spillover $\kappa_4$, the connected four-point function
(fourth joint cumulant) of the centred outcomes $s_i = y_i - \mathbb{E}[y_i]$
(defined in Methods, Eq.~\ref{eq:kappa4})---exactly the part of cluster
behaviour that a Gaussian or pairwise (second-moment) interference model
declares to be zero. Comparing a full-joint
plug-in estimator that uses the empirical four-body structure against a pairwise
maximum-entropy (Ising) model fit to the empirical 1- and 2-body moments only---the
idealized causal-message-passing analogue \cite{noirjean2025exploiting,
bayati2024higher}---at a non-degenerate operating point ($J = 1.8$; maximum
pairwise correlation $0.62$), the true $\kappa_4 = -0.013$, but the pairwise
model is structurally misspecified: it predicts $\kappa_4 = +0.003$---the
\emph{wrong sign} in this constructed cluster---and its absolute error floors
near $0.015$ for every budget from $N = 64$ to $N = 8192$, while the full-joint
estimator converges from $0.006$ to $0.0004$, a $\sim 34\times$ smaller error at
$N = 8192$ (Figure~\ref{fig:results}b). The pairwise model correspondingly
overstates whole-cluster adoption $\Pr(\text{all four adopt})$ at $0.028$ against
the true $0.009$.

The separation is \emph{representational}, not a compute speedup: a pairwise
interference model cannot represent higher-order spillover no matter how much
data it is given, while the entanglement-structured estimand can. Crucially, the
full-joint recovery here is itself classical---a tensor or explicit joint
representation---which is exactly why the realistic near-term substitute is
quantum-inspired rather than quantum. The pilot is illustrative on a deliberately
tiny cluster and makes a conceptual point about representational adequacy, not a
deployable result.

\subsection{A quantum-inspired bandit is runnable but loses to a correctly
specified classical baseline}
\label{sec:res-adaptation}

Adaptive behavioural RCTs---just-in-time adaptive interventions (JITAIs),
micro-randomized trials, sequential nudge designs---increasingly use contextual
bandits and reinforcement learning, and standard estimators applied to
adaptive-trial data can be inconsistent and non-replicable across trial
repetitions even at large $N$ unless the bandit is itself proven \emph{replicable}
\cite{zhang2025replicable}. Quantum bandits carry provable exploration
separations in theory, though established in quantum-state-learning settings
rather than for classical contextual rewards \cite{casale2020quantum,
lumbreras2022multi}; to my knowledge no published RCT has deployed one.

I simulate a six-arm just-in-time nudge intervention for daily physical-activity
behaviour change as a contextual bandit with a three-dimensional context and a
\emph{linear} reward over $T = 1000$ rounds and five seeds, choosing the linear
reward deliberately so that LinUCB---a disjoint per-arm ridge model with
upper-confidence-bound exploration---is the correctly specified, near-optimal
classical baseline. Against it I run a quantum-inspired surrogate: a three-qubit
variational circuit that angle-encodes the context, applies four variational
layers, and reads out six Pauli observables as arm scores under a softmax policy,
trained on-policy by REINFORCE (Methods). The variational policy \emph{does}
learn---its final cumulative regret of $599 \pm 16$ sits $12\%$ below the
random-policy reference of $678 \pm 31$---but LinUCB dominates at $20 \pm 4$,
over $30\times$ lower regret (Figure~\ref{fig:results}c). The conclusion is
robust to tuning: a policy-gradient surrogate cannot match LinUCB's closed-form
ridge updates on a problem built to suit LinUCB. The fair reading is that a
quantum-inspired bandit is \emph{runnable today} but is not competitive with a
well-specified classical baseline; its near-term value is not in beating matched
classical methods.

\section{Discussion}
\label{sec:discussion}

\begin{figure}[t]
\centering
\resizebox{\columnwidth}{!}{%
\begin{tikzpicture}[font=\sffamily, >={Latex[length=2.4mm]},
  dec/.style={diamond, aspect=2.4, draw, align=center, inner sep=0pt,
              text width=3.2cm, fill=blue!8, minimum height=1.4cm},
  term/.style={rectangle, rounded corners, draw, align=center,
               text width=3.4cm, inner sep=4pt, fill=gray!10},
  arr/.style={->, thick, gray}]
\node[term] (start)   at (0,0)      {Pipeline stage with a candidate quantum / quantum-like primitive};
\node[dec]  (d1)      at (0,-2.6)   {Primitive classical-computable today?};
\node[dec]  (d2)      at (-4.7,-5.8){Algorithm mature, only the application missing?};
\node[dec]  (d3)      at (4.7,-5.8) {Beats or reframes the strong classical baseline?};
\node[term] (pilot)   at (-6.9,-9.1){Scope a validated pilot\\[1pt]\emph{(Stages 2, 3)}};
\node[term] (horizon) at (-2.5,-9.1){Research horizon: quantum-inspired surrogate\\[1pt]\emph{(Stage 4)}};
\node[term] (adopt)   at (2.5,-9.1) {Adopt now\\[1pt]\emph{(Stage 1)}};
\node[term] (none)    at (6.9,-9.1) {Keep the classical baseline\\[1pt]\emph{(Stage 5)}};
\draw[arr] (start)--(d1);
\draw[arr] (d1) -- node[above left]{no}  (d2);
\draw[arr] (d1) -- node[above right]{yes} (d3);
\draw[arr] (d2) -- node[left]{yes} (pilot);
\draw[arr] (d2) -- node[right]{no} (horizon);
\draw[arr] (d3) -- node[left]{yes} (adopt);
\draw[arr] (d3) -- node[right]{no} (none);
\end{tikzpicture}}
\caption{The stage-by-stage decision rule as a flowchart. Two questions---whether
the primitive runs on a classical computer today, and whether it beats or
reframes the strong classical baseline---sort each pipeline stage into one of
four verdicts: adopt now, scope a validated pilot, keep the classical baseline,
or treat as a research horizon. Two placements need a gloss. Stage~3's
primitive is classical-computable but has not yet been validated against the
classical baseline in any published RCT, so it routes to the pilot verdict
rather than adoption; Stage~1's primitive is quantum hardware, but it is
consumed as a public beacon service, so it counts as available today.}
\label{fig:f2}
\end{figure}
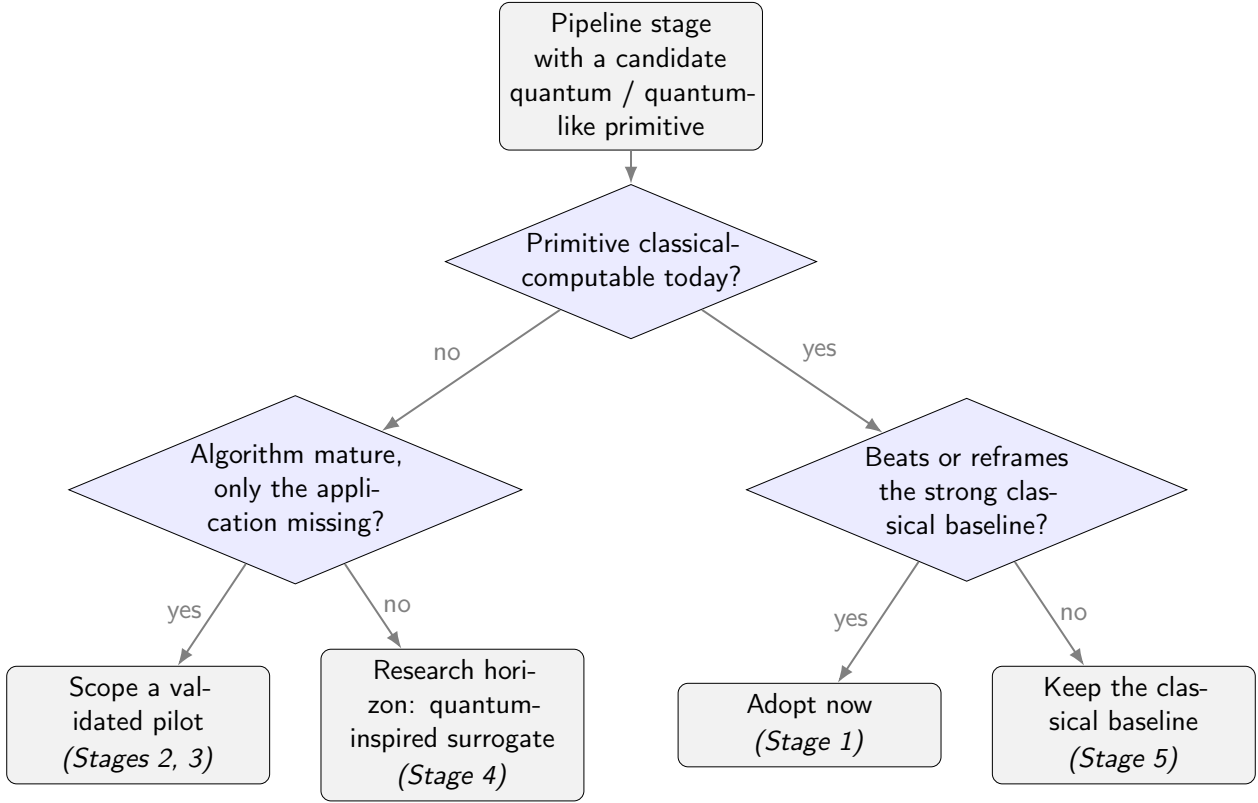

The three pilots return a deliberately mixed verdict---one conditional win
(power), one representational win (interference), one honest loss
(adaptation)---and that mix is the point. The methodologist case for quantum in
behavioural economics is \emph{stage-by-stage}, not all-or-nothing. Two
questions sort each pipeline stage into one of four actions (Figure~\ref{fig:f2}):
does the primitive run on a classical computer today, and does it beat or reframe
the strong classical baseline? Applying that rule across all five stages,
including the two not covered by a pilot, gives the decision rule below.

\subsection{The two stages without a pilot}
\label{sec:design-stages}

\paragraph{Randomization (Stage~1): adopt now.} Pseudo-random number generators
are deterministic given a seed, so the auditability of allocation in
pre-registered nudge RCTs rests on trust in seed storage and is hard to verify ex
post. Device-independent quantum random number generators produce randomness
whose unpredictability is certified by violation of a Bell inequality rather than
by trust in the hardware vendor \cite{pironio2010random, acinmasanes2016certified},
and the NIST / CU Boulder \emph{CURBy} beacon publishes auditable random values
via a blockchain-style Twine protocol \cite{kavuri2025traceable, curby2025beacon}.
A trial that draws a beacon pulse at the pre-registered randomization timestamp,
pre-commits its hash to a public registry before the allocation list exists, and
feeds it to a standard covariate-adaptive minimization routine replaces a
privately held seed with public, pre-committed randomness: a replicator can later
re-fetch the Twine entry, recompute the committed hash, and confirm that the
assignment could not have been chosen---by anyone, including the trial
team---to favour a hypothesis. The upgrade is engineering, not algorithmic, and
the adoption barrier is essentially zero for any team with internet access to the
beacon.

\paragraph{Outcome measurement (Stage~3): ready to pilot.} Survey-based
behavioural RCT outcomes show question-order effects, response-replicability
effects, and ambivalence patterns that violate the assumptions of projective
classical measurement. Khrennikov et al.\ (2025) prove that the
\emph{combination} of question-order and response-replicability effects cannot be
modelled within the projective measurement class, and characterize the
\emph{sharp repeatable non-projective} class as the appropriate generalization
\cite{khrennikov2025quantum, pothosbusemeyer2013can}. This class is
classical-computable: embedding a POVM measurement model in an RCT analysis
pipeline requires no quantum hardware. A pre-registered re-analysis that fits the
non-projective measurement model against a graded-response IRT baseline, with the
question-order-equality violation test as the assumption-falsification gate,
yields paired treatment-effect estimates plus a Bayes factor between the
projective and non-projective measurement classes for the same trial data---all
classical-computable on a laptop. The formalism reframes an existing measurement
problem rather than introducing a new dependency, so Stage~3 is ready to pilot on
the analysis side; no published RCT-integration paper exists.

\subsection{A stage-by-stage decision rule}
\label{sec:decision-rule}

Collecting the five verdicts: adopt certified QRNG now (Stage~1); pilot
quantum-like POVM measurement models now (Stage~3); scope QAE-for-power as the
next methodological project, using enumerated multiverse benchmarks
\cite{veltrigilbert2026multiverse} as the ground-truth target (Stage~2); flag
interference and adaptation as research horizons whose realistic near-term
substitutes are quantum-inspired classical surrogates (Stages~4--5); and insist
on a correctly specified classical baseline at every step. The two pilots that
favour a quantum or quantum-like method do so for different reasons---one
query-complexity (and conditional on state preparation), one
representational---and the one that does not, adaptation, fails against a
baseline built to suit it, which is exactly the discipline Barati et al.\ (2025)
demand \cite{barati2025quantum}.

\subsection{A five-gap research agenda}
\label{sec:agenda}

The decision rule implies five concrete, datable, publishable first papers, in
increasing order of speculative horizon. \textbf{G1:} no paper applies QAE to RCT
sample-size or power calculation; a simulation study comparing classical Monte
Carlo and IQAE \cite{grinko2021iterative} on stylized nudge-RCT power surfaces,
using \cite{veltrigilbert2026multiverse} as the ground-truth benchmark, would
close it. \textbf{G2:} no paper proposes certified QRNG for behavioural-RCT
randomization integrity; a methodological note plus an open-source allocation
tool using the CURBy beacon \cite{curby2025beacon} would. \textbf{G3:} no
published behavioural RCT has deployed a quantum bandit; a pilot JITAI
re-analysis with a quantum-inspired bandit policy benchmarked against the
replicability-preserving classical bandit of \cite{zhang2025replicable} would.
\textbf{G4:} quantum-like decision theory has not been embedded operationally in
the analysis pipeline of a field RCT; a pre-registered re-analysis of a published
multi-question survey-RCT using a non-projective measurement model
\cite{khrennikov2025quantum} versus a classical IRT baseline would. \textbf{G5:}
quantum kernels \cite{schuldkilloran2019quantum, havlicek2019supervised} have not
been integrated into causal-forest heterogeneous-treatment-effect estimation for
behavioural RCTs. The gaps do not map one-to-one onto the five stages
(Figure~\ref{fig:f1}): Stage~4 is the least mature and is framed as a horizon
rather than a single first-paper gap, while G5 is cross-cutting.

\subsection{Limitations}
\label{sec:limits}

I make no measured speedup claims on realistic hardware. The power-pilot ratio of
20--30 times at 64 queries reflects the canonical QAE error bound on a noiseless
state-vector simulator; on noisy intermediate-scale quantum hardware the gain
degrades with circuit depth, and the amplitude-loading step can erase the
advantage for an unstructured distribution. The interference and adaptation
pilots are deliberately small---a four-qubit cluster and a six-arm bandit---and
are illustrative of representational adequacy and of a fair-but-losing comparison
respectively, not deployable results. The power anchor uses a 36-pipeline grid
small enough that a classical analyst would simply enumerate it; the QAE
advantage matters only when the multiverse grid scales beyond exhaustion. I do
not address the metaphysics of quantum cognition: the \emph{quantum-like} hedge
is maintained throughout, and the methodologist case does not depend on quantum
effects in the brain. I restrict the substantive anchor to behavioural economics
and nudge field experiments; psychology, public-health, and digital-behavioural
RCT readers should find the framing largely transferable but not specialized.

\section*{Acknowledgements}
The author thanks Joshua Gilbert for the Veltri \& Gilbert (2026) multiverse RCT
replication archive used as the testbed for the power pilot. The author declares
co-authorship of the testbed paper.

\section*{Statements and Declarations}

\paragraph{Funding.} The author received no specific grant from any funding
agency in the public, commercial, or not-for-profit sectors for this work.

\paragraph{Competing interests.} The author declares no competing interests. The
Stage~2 anchor relies solely on publicly available data (see Data and code
availability); G.A.V.'s co-authorship of the testbed paper, Veltri \& Gilbert
(2026), is disclosed in the Acknowledgements, and no claim is made about the
substantive RCTs in that dataset.

\paragraph{Ethics approval.} Not applicable. This study analyses only simulated
data and publicly available secondary data; no new human-subjects data were
collected.

\paragraph{Consent.} Not applicable (no human participants were involved in this
study).

\paragraph{Data and code availability.} The simulation scripts and pilot records
for all pilots (Stage~2 amplitude estimation, synthetic and real-data; Stage~4
entangled-state spillover; Stage~5 quantum-inspired bandit) are openly available
on the Open Science Framework \cite{quantumbepilots2026} (MIT license). The
scripts are written in PennyLane and run on the CPU \texttt{lightning.qubit} /
\texttt{default.qubit} backends; each pilot ships with a seeded configuration, a
JSON results file, and an environment snapshot for reproducibility. The Stage~2
real-data anchor uses only publicly available data: outcomes from the openly
accessible Item Response Warehouse \cite{domingue2025irw}, enumerated into a
preprocessing multiverse by Veltri \& Gilbert (2026)
\cite{veltrigilbert2026multiverse}.

\paragraph{Author contributions.} G.A.V.\ designed the study, ran the
simulations, and wrote the manuscript.

\bibliography{citations}

\end{document}